\documentclass[conference]{IEEEtran}
\IEEEoverridecommandlockouts

\usepackage[normalem]{ulem}
\usepackage[nottoc]{tocbibind}
\usepackage{multicol}
\usepackage{cite}
\usepackage{amsmath,amssymb,amsfonts}
\usepackage[ruled, vlined, linesnumbered]{algorithm2e}
\usepackage{graphicx}
\usepackage{subfigure}
\usepackage{comment}
\usepackage{textcomp}
\usepackage{xcolor}
\usepackage{enumitem}
\usepackage{algorithm2e}

\usepackage{caption}
    \bstctlcite{IEEEexample:BSTcontrol}

\begin{document}

\title{\huge DNN-based Denial of Quality of Service Attack on Software-defined Hybrid Edge-Cloud Systems\thanks{This material is based upon work supported by the National Science Foundation under Award Number: CNS-1943338. Any opinions, findings, and conclusions or recommendations expressed in this publication are those of the author(s) and do not necessarily reflect the views of the National Science Foundation.
}}

\author{
Minh Nguyen\IEEEauthorrefmark{1}, Jacob Gately\IEEEauthorrefmark{2}, Swati Kar\IEEEauthorrefmark{2}, Soumyabrata Dey\IEEEauthorrefmark{2}, Saptarshi Debroy\IEEEauthorrefmark{1}\\
\IEEEauthorblockA{\IEEEauthorrefmark{1} Computer Science, City University of New York, New York, NY 10065}
	\IEEEauthorblockA{\IEEEauthorrefmark{2} Computer Science, Clarkson University, Potsdam, NY 13699}
Email:{\{\textit{mnguyen@gradcenter.cuny.edu, \{gatelyjm, kars, sdey\}@clarkson.edu, saptarshi.debroy@hunter.cuny.edu}\}}}
\maketitle

\begin{abstract}
   In order to satisfy diverse quality-of-service (QoS) requirements of complex real-time video applications, civilian and tactical use cases are employing software-defined hybrid edge-cloud systems. One of the primary QoS requirements of such applications is ultra-low end-to-end latency for video applications that necessitates rapid frame transfer between end-devices and edge servers using software-defined networking (SDN). Failing to guarantee such strict requirements leads to quality degradation of video applications and subsequently mission failure. In this paper, we show how a collaborative group of attackers can exploit SDN's control communications to launch Denial of Quality of Service (DQoS) attack that artificially increases end-to-end latency of video frames and yet evades detection. In particular, we show how Deep Neural Network (DNN) model training on all or partial network state information can help predict network packet drop rates with reasonable accuracy. We also show how such predictions can help design an attack model that can inflict just the right amount of added latency to the end-to-end video processing that is enough to cause considerable QoS degradation but not too much to raise suspicion. We use a realistic edge-cloud testbed on GENI platform for training data collection and demonstration of high model accuracy and attack success rate.
\end{abstract}


\section{Introduction}
In order to satisfy diverse quality-of-service (QoS) requirements of complex real-time applications (e.g., video processing, 3D reconstruction, AR/VR), civilian and tactical use cases are starting to employ software-defined hybrid edge-cloud systems (a.k.a. Software-defined Wide Area Network)
\cite{sd-wan1,sd-wan2}.
The edge sites are primarily responsible for processing real-time jobs (at the edge servers) in order to satisfy the end-to-end latency requirements of such applications. Connectivity to the cloud is essential for: i) processing significantly intensive computation jobs offloaded by edge sites and ii) running the centralized Software-defined (SDN)~\cite{sdn} controller that manages edge-cloud resources through OpenFlow~\cite{openflow} based control communication with the edge sites.


Unlike traditional distributed networking, SDN uses a centralized approach where a programmable control plane (hosted by a SDN controller) dictates routing through fast, simple, and commodity routers/switches implementing generalized data-plane forwarding in hardware. In this approach, OpenFlow API is used by the switches to request forwarding rules that are computed by the central controller in order to route video packets from camera-enabled end-devices to edge servers for processing~\cite{openflow}. In hybrid edge cloud systems, SDN controllers are often hosted at remote cloud data-centers (instead of one of the edge sites) in order to give the controller global visibility of the entire environment~\cite{sd-wan1}. However, this adds Internet scale delays ($>$100 ms) to the end-to-end latency of video processing for control communication between the switch and the controller. If persistent, Such delays can severely impact the quality of service (QoS) of real time video applications resulting in failure of involved missions~\cite{real-time}.   


In this paper, we propose a stealthy attack that can cause such persistent Denial of QoS (DQoS) for edge-cloud supported real-time video applications. Unlike traditional Denial of Service (DoS) attacks that aims to cripple normal operations of a system by flooding the target at high intensity, our proposed stealthy DQoS aims to increase the end-to-end video packet delivery delay just enough to violate the application QoS requirement, yet stay undetected. Our proposed DQoS attack uses deep neural networks (DNN) to artificially cause frequent control communication exchange between the routers (at the edge) and the cloud-native SDN controller, thereby increasing the end-to-end delivery latency of video packets from end-devices to the edge servers. We show that a small group of collaborative attackers infiltrating the system and monitoring all or some of the key system/network information (with/without noise) can effectively (i.e., with very low error rate) train a DNN model to predict key network metrics (e.g., average packet drops at strategic routers) for different attack intensities. We show how such prediction can help the attackers in generating ideal attack intensity that does not raise any system alarms (i.e., being stealthy) and yet can significantly increase the video packet latency. It is to be noted that the focus of this paper is not to investigate how the attackers can monitor such network parameters, rather how such parameters can be exploited (if compromised) to significantly impact the application QoS.
We evaluate the performance of the proposed attack model on a softwarized egd-cloud testbed implemented on GENI framework~\cite{geni}. The results demonstrate that the proposed attack model can increase the end-to-end latency (from end-devices to edge severs) by $\sim$3x without increasing the average packet drop rate beyond the acceptable range.


\section{System Model and Problem Evidence Analysis}
\begin{figure}[t]
\centering
  \includegraphics[width=\linewidth]{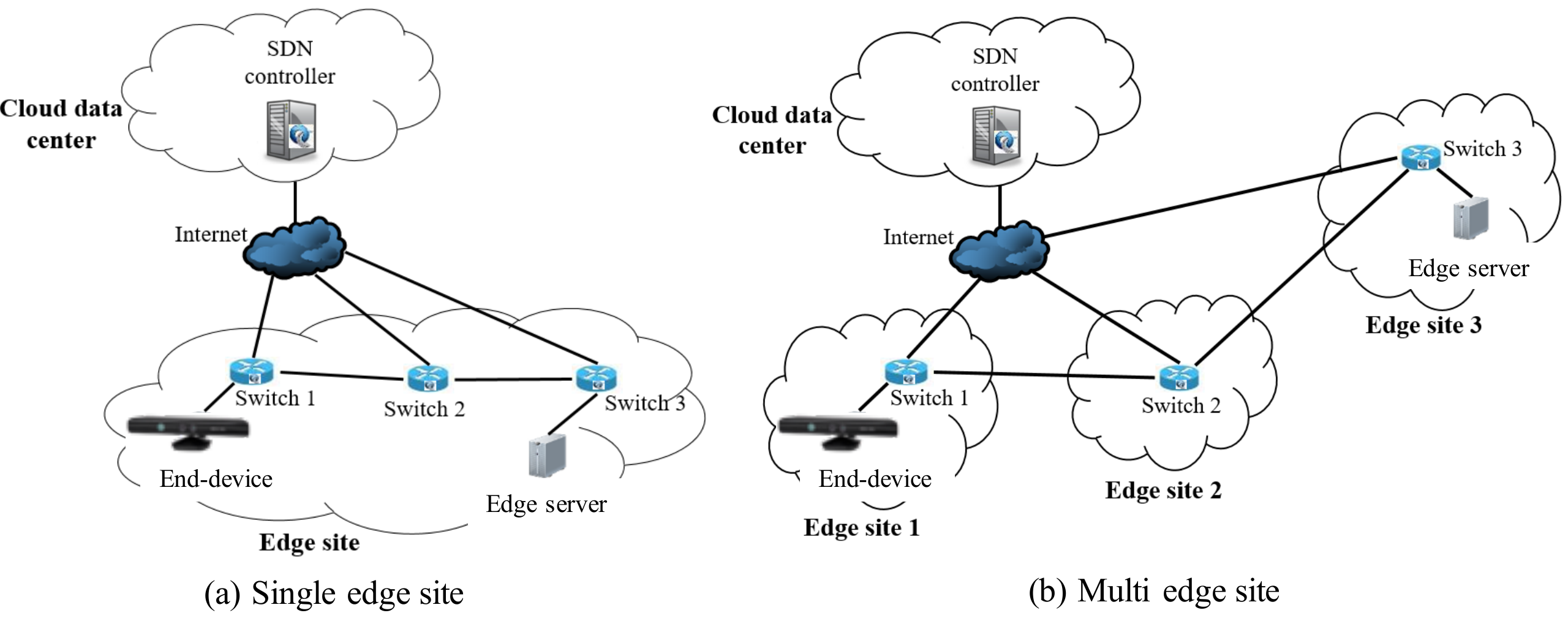}
\caption{\footnotesize{Single vs. multi-site software-defined hybrid edge-cloud models}}
\label{Fig:system model}
\vspace{-0.3in}
\end{figure}

As shown in Fig.~\ref{Fig:system model}, software-defined hybrid edge-cloud systems that support real-time video processing applications involve single or multiple edge site(s) consisting of end-devices that capture videos and send the video frames to an edge server (in the same or different edge site) for processing through OpenFlow switches. The SDN-enabled OpenFlow switches route packets containing video frames using their flow tables that are populated by the cloud-hosted SDN controller. 
Typically, for any incoming packet, the OpenFlow switch consults its flow table to check if any existing flow rule applies for that packet (based on packet contents, e.g., source IP, destination IP etc.) in order to forward it.
However, unlike traditional networking, if no such flow rules exist in the table (an event called {\em Table miss}), the switch asks the SDN controller for new rules (in form of \texttt{Packet-In} messages), in response to which the controller computes and pushes new flow rules (in form of \texttt{Packet-Out} and \texttt{Flow-Mod} messages) to the requesting switch. At the switch, these rules are then added to the flow table for forwarding all packets that match with those flow rules~\cite{openflow-rules}.

The size of flow tables in OpenFlow switches can vary within a wide range. However for a particular network, the size typically depends on the switch's hardware capacity and system administrator's implementation needs. Based on the size of the flow table and with each new flow addition, at one point a flow table gets fully occupied and beyond which no new flow can be added. To work around this, OpenFlow protocol uses two data structures, viz., \texttt{Hard Timeout} and \texttt{Idle Timeout} that define the periodicity of erasing older flow rules from the table in order to make room for new flows. Again, the values of such timeouts depend on system implementation. System implementation also dictates parameters and metrics that are used by the SDN controller to monitor suspicious behavior and anomalies. There is a range of such metrics and parameters used by different systems based on the system realities and objectives, e.g., packet drop rate, bandwidth utilization, and switch/router buffer overflow status to name a few. In this work, we assume that the controller monitors the packet drop rate at strategic switches as an indicator for network anomalies and sustained drop rates beyond the statistical long term average is considered as an indicator of suspicious behavior. 

In traditional data center implementation of SDN, any such table miss only causes a negligible amount of delay for round-trip communication between the switch and controller. However for software-defined edge-cloud implementations, any table miss add Internet scale delay to the end-to-end latency of video frames and if recurrent can cause Denial of QoS. We use the CloudLab~\cite{cloudlab} platform to perform evidential experiments to demonstrate the feasibility of table misses initiated end-to-end latency increase of video frames. As shown in Fig.~\ref{Fig:system model}, we use two different topologies where the end-devices and edge servers are in the same and different edge sites. Fig.~\ref{Fig:system model}(a) shows an edge-cloud system where video frame source and destination within the same edge site (located in Clemson aggregate within CloudLab platform) are connected through 3 OpenFlow switches that are again interconnected through high bandwidth dedicated Layer 2 connection. Fig.~\ref{Fig:system model}(b) shows a multi-site scenario where the edge sites (located in Clemson, UWisconsin, and Utah aggregates) are interconnected via Layer 3 Internet. In both scenarios, the edge sites are connected via Internet to the data center hosting the SDN controller (located in UMass aggregate).

\begin{table}[htb]
\caption{Avg. latency for table misses at different switches along the route}
\label{Table:latency-comp}
\centering
\begin{footnotesize}
\begin{tabular}{|c||c|c|}
\hline
\bfseries Scenarios & \bfseries 
\begin{tabular}{@{}c@{}}Same\\ 
site\end{tabular}
& \bfseries 
\begin{tabular}{@{}c@{}}Different\\ sites \end{tabular}
\\
\hline\hline
\begin{tabular}{@{}c@{}}Avg. latency for no table miss \end{tabular} & 10 ms & 66 ms\\
\hline
\begin{tabular}{@{}c@{}}Avg. latency for table miss at S1 only \end{tabular} & 78 ms & 144 ms\\
\hline
\begin{tabular}{@{}c@{}}Avg. latency  for table miss at S1 and S2\end{tabular} & 144 ms & 213 ms\\
\hline
\begin{tabular}{@{}c@{}}Avg. latency for table miss at all 3 switches \end{tabular} & 212 ms & 280 ms\\
\hline
\end{tabular}
\vspace*{-0.3in}
\end{footnotesize}
\end{table}

Table~\ref{Table:latency-comp} compares average latency for table misses at different switches along the route from the end-device to the edge server for single and multi-site edge-cloud scenarios. The results show that for both cases, each additional table miss along the route adds upto $\sim$70 ms of roundtrip (from switch to controller and back) delay to the end-to-end latency from end-device to edge server. It is important to keep in mind unlike this controller setup, in real production edge-cloud network the roundtrip delay for each table miss will be much higher ($>$100 ms) due to much higher traffic load at the switch and controller. The overall results clearly demonstrates the attacker objective, i.e., {\em making sure that frequent table misses occur at as many switches as possible along the route without increasing the mean packet drop rate at strategic switches.}

\section{Attack Model Design and Evaluation}
\subsection{Attack model approach}
\label{Sec:approach}
In order to achieve the above objective, we take a data-driven approach where we assume that a group of collaborative attackers have infiltrated the edge sites by connecting to the end switches/base stations where typically other end-devices would connect through address spoofing. Typically in edge-cloud systems, such end-connections would be wireless that opens the door for many snooping and infiltration vulnerabilities. We assume that such vulnerabilities aid the attackers to gain partial or complete view of the overall system and its parameters. Based on such partial or complete view, the attackers can monitor system behavior by injecting unsuspecting traffic into the network as part of reconnaissance. Later we will show the attack success is correlated with gaining such partial or complete view and their ability to observe different parameters. For the reconnaissance, we assume that the attackers have a wide degree of freedom (e.g., spoofed IP address, MAC address, port number) in terms of parameters for false packet injection. Based on the observed network parameters for a certain false packet injection intensity, the attackers (enabled with enough processing capability) train a DNN that can predict mean packet drop rates at strategic switches for a certain packet injection intensity. This prediction is then used to design a greedy algorithm to generate ideal false packet injection rate that can significantly increase end-to-end latency of regular video traffic by triggering frequent table misses at multiple switches along the route~\cite{zhang2018control} without increasing mean packet drop rates in important switches.

\subsection{Testbed design for training data collection}
\begin{figure}[t]
\centering
  \includegraphics[width=0.7\linewidth]{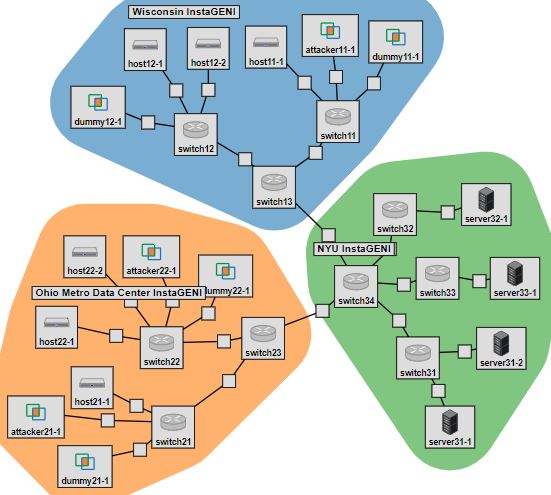}
\caption{\footnotesize{An experimental edge-cloud testbed implementation on GENI platform}}
\label{Fig:topology}
\vspace{-0.2in}
\end{figure}

In order to collect the dataset required for DNN training, we generate synthetic data using a realistic edge-cloud testbed implemented on GENI platform~\cite{geni} (as shown in Fig.~\ref{Fig:topology}) as real data for such next generation systems is not readily available. NSF-supported GENI platform is an educational cloud environment that allows registered users to create distributed testbeds as collection of virtual machines (VMs) in federated sites across the continental US and abroad for running futuristic experiments. As can be seen in Fig. \ref{Fig:topology}, our softwarized hybrid edge-cloud testbed implements three edge sites: one at NYU InstaGENI that hosts all the edge servers for video data processing and two at Wisconsin InstaGENI and Ohio Metro Data Center InstaGENI that primarily host end-devices that generate video traffic for processing. The host edge sites together have 6 hosts (as end-devices) that send serialized video frames to the 4 servers for processing. 
Each host's video transmission follows an ON-OFF model where during the ON phase the host sends frames and stays idle during the OFF phase. Length of both ON and OFF phases are chosen randomly within the ranges of $10-15$ and $0-5$ minutes respectively. The ranges correspond to typical operating times of off-the-shelf camera enabled drones/robots between recharges. 
The host to server mapping follows a fair allocation method where servers are selected for a particular session in order to keep balanced traffic loads among servers.


The host edge sites also consist of collaborative attackers, as well as dummy VMs that help generate artificial traffic that mimic the traffic load in realistic edge-cloud environments. Specifically, dummy VMs generate artificial traffic to keep the mean packet drop rate around 0\% with $+3$\% maximum allowance on \textit{switch34}, which is the most important switch in the system for being the entry point to the server site and a point of potential traffic bottlenecks. The host edge sites are connected to the server edge site through a collection of Open Virtual Switches that act as programmable switches and create the softwarized network fabric. The cloud-hosted SDN controller (not shown in this figure) connects to the switches via Internet and implements Floodlight for control plane. For the switches' configuration, the flow table sizes are set to 1000 flow entries, \texttt{Idle Timeout} is set to 5 seconds, and \texttt{Hard Timeout} to $\infty$, commensurate with a system of this proportion. The bandwidth for the connections between edge switches (e.g., between \textit{switch34} and \textit{switch13} or \textit{switch34} and \textit{switch23}) is set to 100 Mbps while for all other connections it is set to 50 Mbps.

\subsection{Data collection and preparation}
We collect historical network state versus packet drop rates data for different attack (i.e., false packet generation) intensities using a  Python based tool that is able to forge packets~\cite{attack-tool}. The network state is defined as a vector of comprehensive network parameters, viz., packet drop rates and flow table sizes at different switches, end-to-end latency of video frames, and bandwidth utilization at different links etc.
In our experiments towards data collection, each attacker $a_i$ can generate packets at rate $\alpha_i^t\times R$ at any time instance $t$, where $\alpha_i^t \in [0, 1]$ represents the fraction of maximum attack rate $R$.  
In this experiment, $R$ for each attacker is set to 100 packets per second with each packet payload is designed to cause table-miss (as explained in Section~\ref{Sec:approach}).
Given the testbed architecture and experimental setup, the network state parameters change with attack state, i.e., attackers' packet generation rates. At every $t=10$ seconds, we randomly select an attack vector $\{ \alpha_i^t \}_{i=1}^K$ ($K$ is the number of attackers) for all attackers and observe the corresponding network state changes. A measurement snapshot consists of the network state at current time $t$ (current state), attack rates $\{ \alpha_i^t \}_{i=1}^K$, and the changed network state (changed state) from $t+5$ seconds to $t+10$ seconds. Here the $+5$ seconds delay between generating a new attack state and capturing the corresponding network state ensures strong causality. The network state information collection between $t+5$ to $t+10$ takes place at a granularity of $100$ ms. 
The total experiment is run for over 3 hours generating around $1200$ snapshots.







\begin{figure*}[t]
    \centering
    \subfigure[\footnotesize{Switch34 with Model 1}] {
        \includegraphics[width=0.8\columnwidth]{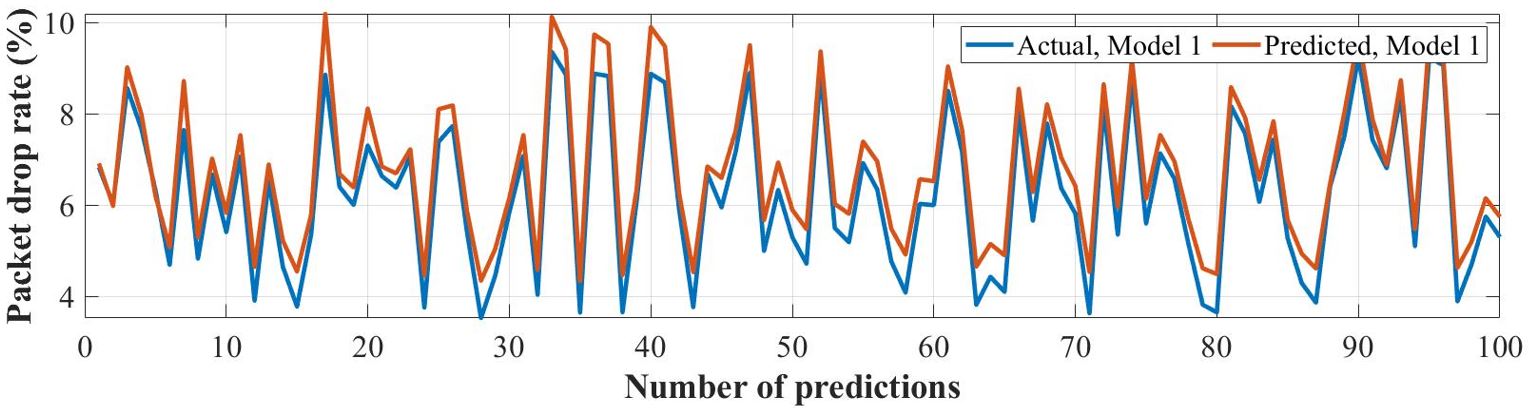}
        \label{Fig:switch34-model1}
    }
    \subfigure[\footnotesize{Switch34 with Model 7}] { 
        \includegraphics[width=0.8\columnwidth]{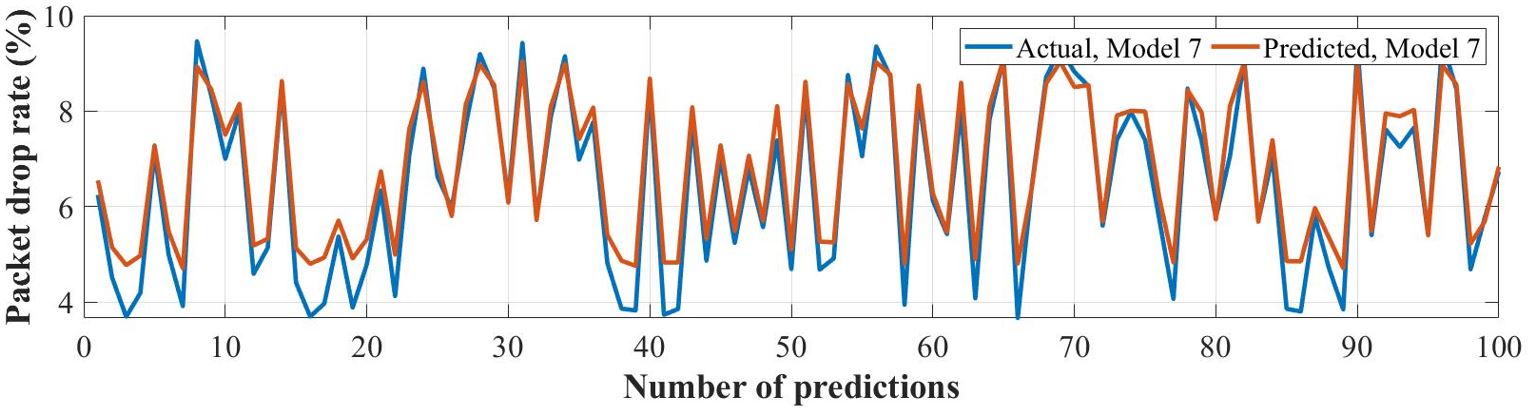}
        \label{Fig:switch34-model7}
    }\\
    \subfigure[\footnotesize{Switch21 with Model 4}] { 
        \includegraphics[width=0.8\columnwidth]{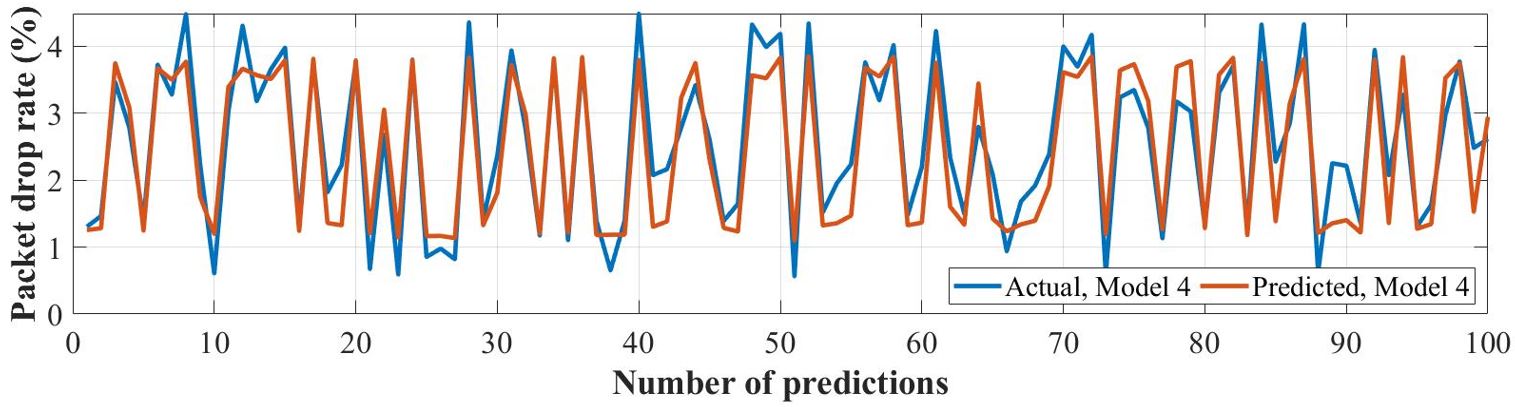}
        \label{Fig:switch21-model4}
    }
    \subfigure[\footnotesize{Switch34 and Switch21 with Model 10}] { 
        \includegraphics[width=0.8\columnwidth]{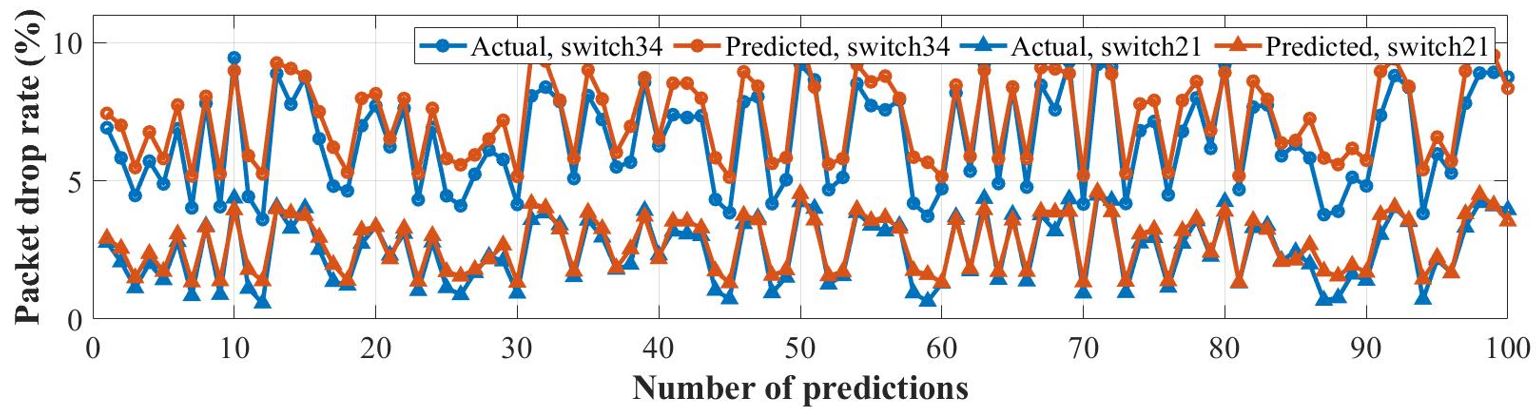}
        \label{Fig:model10-2switches}
    }
    \caption{Packet drop rate prediction results for different switches using different models}     
    \label{test_results}
    \vspace*{-10pt}
\end{figure*}

\subsection{DNN model design and training} 
Using the collected data, we train a DNN model that can predict the packet drop rates at strategic locations, e.g., {\em switch34} from Fig.~\ref{Fig:topology}. Later, we will show how this aids the attackers to adjust the attack parameters that is ideal to trigger enough table misses at different switches to cause maximum latency increase but limit the drop rates within normal network bounds. 
To this end, we design a DNN that is simple in architecture but highly efficient in predicting the packet drop rates. The network architecture is defined in Table \ref{Table:DNN_architecture}. The DNN model is optimized on the training dataset to be able to accurately predict the drop rates. Each iteration of DNN training is performed on a batch of $5$ snapshots where each snapshot consists of current network state, $\{ \alpha_i^t \}_{i=1}^K$, and the corresponding changed network state. The input to the DNN is a pair of current state and $\{ \alpha_i^t \}_{i=1}^K$ that is feed forwarded through the DNN layers to predict the changed drop rates. A batch of such inputs generates the predicted drop rate sets that are compared with the corresponding actual drop rates of the changed states. A mean square loss function is used to compute the average prediction error per batch which is then back propagated from output to input layers resulting appropriate changes of network parameters towards loss minimization. The \textit{adam} optimizer is used for network training. The network is trained for a maximum of 50 epochs, but an early stopping mechanism with patience 5 is used to prevent overfitting and to help minimize loss.

Depending on the extent of system infiltration level, the attackers may have access to complete or partial network state. at the same time, attackers' visibility of the network state may or may not be accurate. With this in the mind, we trained 10 variations of the DNN model. We divide network state parameters into $5$ categories: (1) time difference between successive frames for each host to server connection, (2) packet drop rates at each switch, (3) flow table size at each switch, (4) bandwidth utilization between each pair of connected switches, and (5) number of waiting frames at each switch. All the models use $\{ \alpha_i^t \}_{i=1}^K$ and a subset of network state categories. Model 1 uses all categories. Model 2 uses categories 1, 3, 4, and 5. Model 3 uses categories 1 and 3. Model 4 uses categories 1 and 4. Model 5 used categories 1 and 5. Finally, Model 6 through 10 use inputs similar to Model 1 through 5 but the network state values are polluted with a Gaussian noise with standard deviation of 0.3.


\begin{table}[t]
\begin{center}
\caption{Summary of DNN model architecture.}
\label{Table:DNN_architecture}
\begin{tabular}{ |c|c|c|c|c| } 
 \hline
 Layer & Type & Nodes & Act. Function & Rate \\ 
 \hline
 \hline
 1 & Dense & 20 & relu & - \\ 
 \hline
 2 & Dropout & - & - & 0.15 \\ 
 \hline
 3 & Dense & 20 & relu & - \\ 
 \hline
 4 & Dropout & - & - & 0.15 \\ 
 \hline
 5 & Dense & 20 & relu & - \\ 
 \hline
 6 & Dropout & - & - & 0.15 \\ 
 \hline
 7 & Dense & 10 & - & - \\ 
 \hline
\end{tabular}
\end{center}
\vspace{-0.4in}
\end{table}

\subsection{Performance evaluation of DNN training}
Fig.~\ref{test_results} shows the accuracy of different prediction models. 
In order to produce these results, we run a new set of experiments for $1000+$ seconds where $\alpha_i^t$ are changed in every 10 seconds. The attack rates are randomized, however their range is chosen in order to keep the packet drop rates between 3\% to 10\%. During an attack interval, the $+5$ seconds of delay between the new attack state and corresponding network state is maintained similar to test data collection method. For the performance evaluation, we predict the current packet drop rates of all switches using the current attack state and the previous network state  and compare them with the actual current packet drop rates. We repeat the same experiment for all the models. 
Fig. \ref{Fig:switch34-model1} and \ref{Fig:switch34-model7} illustrate the actual and predicted packet drop rates of \textit{switch34} with Model 1 
and Model 7 
respectively. 
Meanwhile, Fig. \ref{Fig:switch21-model4} compares the actual and predicted but for \textit{switch21} with Model 4. 
Finally, Fig. \ref{Fig:model10-2switches} compares drop rates of \textit{switch34} and \textit{switch21} with Model 10. Overall, all the results demonstrate high accuracy of the proposed DNN based packet drop predictions.

\subsection{DQoS attack algorithm design and performance}
We use a simple greedy algorithmic approach (as shown in Algo. 1) for the attackers to compute ideal attack rates in order to maintain the average packet drop rate of \textit{switch34} (the most important switch) between it's statistical mean of ($m\pm k$)\% as it is assumed that such packet drop rate does not trigger any system alarms. Due to the high accuracy of the DNN based prediction of packet drop rates, it is used by the attackers to adjust the ideal attack rate $\alpha_i$. As the network conditions are dynamic, the algorithm is run periodically in order to adjust the attack rates based on current network conditions.
Table~\ref{Table:Latency_degradation} shows the end-to-end latency degradation of video packets under the ideal attack intensity computed by Algo. 1. We observe that for all possible maximum allowed drop rate scenarios for {\em switch34}, the mean latency under attack is $\sim$3x worse than the latency without attack. Upon investigation, we find that the attack is able to successfully cause frequent table misses at all the switches along the route. Such tables misses are caused by the false packets rapidly: (1) populating the flow tables with useless flow rules and (2) replacing old flow rules that are being vacated by \texttt{Hard Timeout} and \texttt{Idle Timeout} stipulations. It is to be noted that $\sim$3x latency increase is due to table misses at only 4 switches along the path. In real edge-cloud implementations, the number of such switches will be higher, resulting in greater degradation.   


\begin{algorithm}[t]
\scriptsize
\caption{DQoS attack rate adjustment algorithm}
\KwIn{Attack rates $K$ attackers $\{ \alpha_i\}_{i=1}^K$; Predicted packet drop rate of \textit{switch34} $dr^{p}_{34}$ using trained DNN model; Statistical actual mean packet drop rate $dr^{a}_J$ ($m\pm k$)\% of Switch $J$; Attack rate increment $p$; Attack rate decrement $q$}
\KwOut{Adjusted attack rates to maintain $dr^{p}_{34}$ between ($m\pm k$)\%}
\While {$dr^{p}_{J} < m\%$}{
    \For {all $i\leq K$}{
        $\alpha_i$ += $p$\;
    }
    Compute $dr^{p}_{J}$;\\
    \While {$dr^{p}_{J} > (m+k)\%$}{
        \For {all $i\leq K$}{
            $\alpha_i$ -= $q$\;
        }
        Compute $dr^{p}_{J}$;
    }
}
\KwRet $\{ \alpha_i\}_{i=1}^K$;
\label{algo:greedy}
\end{algorithm}

\begin{table}[t]
\begin{center}
\caption{End-to-end latency degradation under DQoS attack}
\label{Table:Latency_degradation}
\begin{tabular}{ |c|c|c|c| } 
 \hline
 \begin{tabular}{@{}c@{}}Actual mean\\ drop rate\end{tabular} &  \begin{tabular}{@{}c@{}} Max. allowed\\drop rate \end{tabular}& \begin{tabular}{@{}c@{}} Mean latency\\ without attack \end{tabular}&  \begin{tabular}{@{}c@{}} Mean latency\\ under attack\end{tabular}\\ 
 \hline
 \hline
 0\% & +1\% & 55 ms & 133 ms\\
 \hline
 0\% & +2\% & 58 ms & 154 ms\\
  \hline
 0\% & +3\% & 60 ms & 161 ms\\
 \hline
\end{tabular}
\end{center}
\vspace{-0.3in}
\end{table}

\section{Conclusions}
In this paper, we showed a proof-of-concept of DNN training enabled stealthy collaborative attack to significantly degrade QoS of edge-cloud system hosted real-time video processing applications. As part of future work, we plan to explore the means for such attackers to monitor the network parameters for successful DNN training. At the same time, we will explore how systems can exploit AI based techniques to thwart such stealthy attacks.

\scriptsize
\bibliographystyle{IEEEtran}
\bibliography{sdn}

\end{document}